\documentclass[aip,graphicx, reprint]{revtex4-1}

\draft 
\usepackage{graphicx}
\graphicspath{{img/}}
\usepackage{xcolor}
\usepackage{amssymb}
\usepackage{booktabs}
\usepackage{tabularx}
\usepackage{booktabs}
\usepackage{times}
\usepackage{array}
\usepackage{natbib}
\usepackage{color}
\begin{document}


\title{ Density Functional Theory based Study of Chlorine Doped WS$_2$-metal Interface } 



\author{Anuja Chanana and Santanu Mahapatra}
\email[]{santanu@dese.iisc.ernet.in}
\affiliation{Nanoscale Device Research Laboratory, Department Of Electronic Systems Engineering, Indian Institute of Science, Bengaluru, Karnataka 560012, India }


\date{\today}

\begin{abstract}
Investigation of a TMD-metal interface is essential for the effective functioning of monolayer TMD based field effect transistors (FETs). In this work, we employ Density Functional Theory (DFT) calculations to analyze the modulation of the electronic structure of monolayer WS$_2$ with chlorine doping and the relative changes in the contact properties when interfaced with gold and palladium.  We initially examine the atomic and electronic structures of pure and doped monolayer WS$_2$ supercell and explore the formation of mid gap states with band splitting near the conduction band edge. Further we analyze the contact nature of the pure supercell with Au and Pd. We find that while Au is physiosorped and forms n-type contact, Pd is chemisorped and forms p-type contact with a higher valence electron density. Next, we study the interface formed between the Cl-doped supercell and metals and observe a reduction in the Schottky barrier height (SBH) in comparison to the pure supercell. This reduction found is higher for Pd in comparison to Au which is further validated by examining the charge transfer occurring at the interface. Our study confirms that Cl doping is an efficient mechanism to reduce the n-SBH for both Au and Pd which form different types of contact with WS$_2$.
\end{abstract}


\maketitle 

After the fabrication of FET using monolayer MoS$_2$,\cite{radisavljevic2011single} 2D layered transition metal dichalcogenides(TMDs) have garnered enormous attention in the electron devices community. Apart from MoS$_2$, other TMDs as such WS$_2$\cite{:/content/aip/journal/apl/101/1/10.1063/1.4732522, doi:10.1021/nn505253p}, WSe$_2$\cite{:/content/aip/journal/apl/103/10/10.1063/1.4820408}, MoSe$_2$\cite{:/content/aip/journal/apl/101/22/10.1063/1.4768218} and MoTe$_2$\cite{doi:10.1021/nn501013c} are also explored as channel material for FET's. In the absence of efficient doping techniques, these transistors exhibit SBH at source/drain contact which leads to low ON current. Tremendous efforts are dedicated to reduce the contact resistance at TMD-metal interface by employing different techniques both theoretically and experimentally.\cite{leong2014low, 7116528, doi:10.1021/nl303583v, :/content/aip/journal/apl/104/9/10.1063/1.4866340, laskar2014p, doi:10.1021/nl503251h} However, most of these efforts are focused towards MoS$_2$ and WSe$_2$ and a minimum study is devoted to other TMD's-metal contact interfaces.  A study using ballistic MOSFET model reveals that WS$_2$ outperforms all other TMD's\cite{5959195}. Experimental reports of WS$_2$ device fabrication \citep{:/content/aip/journal/apl/101/1/10.1063/1.4732522} and Cl doping technique for reducing WS$_2$-metal contact resistance are also reported\cite{doi:10.1021/nl502603d}. However, theoretical investigations of WS$_2$ metal contact interface using first principles is still lacking in the literature. Since first principles are extensively used to analyze the graphene-metal\cite{PhysRevB.79.195425, :/content/aip/journal/jap/108/12/10.1063/1.3524232}, MoS$_2$-metal \cite{PhysRevLett.108.156802} and WSe$_2$-metal\cite{PhysRevX.4.031005}, it is believed that it will efficiently describe the contact nature with other TMD's as well. For WS$_2$, chlorine doping, which is done by replacing sulfur atoms is the first ever method demonstrated experimentally to reduce the WS$_2$-contact resistance, exhibiting both high drain current and field-effect mobility.\cite{doi:10.1021/nl502603d} In this work we employ density function theory (DFT) to study the electronic structure of the interface between WS$_2$ and one physiosorped (Au) and one chemisorped (Pd) metal. We then examine the effect of chlorine doping by substituting sulfur atoms to address the SBH mitigation at WS$_2$-metal interface. 

We start by analyzing the electronic dispersion characteristics of pure and Cl-doped WS$_2$ and explore the shifts in the energy bands with respect to Fermi level (E$_f$) and further analyze the mid gap states formed near the conduction band (CB) edge. To preserve the stability in the 5x5 WS$_2$ supercell, the number of chlorine dopants substituting the sulfur atoms is kept one. We perform the formation energy calculations (E$_{Form}$) to find the stability of the chlorine doped structure. Next, this supercell is interfaced with $<$111$>$ cleaved surfaces of Au and Pd respectively and the contact nature is studied. The Schottky barrier height (SBH) is evaluated using the projected bandstructure and the density of states. The charge transfer across the interface is analyzed using valence electron density and charge density difference. Electron localization function (ELF) is employed to study the localization of electrons at the WS$_2$-metal interface. A thorough examination of all the above analysis leads to a conclusion that n-type doping of WS$_2$ using chlorine as an effective substitute, results in lowering of n-SBH at the WS$_2$-metal interface.

We carry out the DFT simulations employing Atomistix Tool Kit (ATK)\cite{QumWS} with Local Density Approximation (LDA)-Perdew Zung (PZ)\cite{PhysRevB.23.5048} as the exchange correlation to investigate the WS$_2$ (pure and doped) - metal (Au and Pd) interfaces. The pseudopotentials of various elements (Tungsten, Sulfur, Chlorine, Gold and Palladium) as generated by Hartwingster–-Goedecker–-Hutter \cite{PhysRevB.58.3641} based on the concept of fully relativistic all electron calculation are used for the present study. The orbital contribution in Tier 4 basis set is found to be appropriate for generating the band gap of monolayer WS$_2$. It is worth noting that in transition metals (e.g W and Pd) the overlap between core and valence electrons is substantial. The commoonly used double zeta polarized (DZP) basis function generated using FHI pseudopotential does not consider the semicore electrons and might not yield results of considerable accuracy. On the other hand HGH pseudopotentials include semi core electrons and thus are expected to produce the electronic structure of the TMD-metal interface precisely.  However, their usage significantly enhances the computational budget since they require a larger number of basis functions as compared to DZP. The density mesh cut off is 75 Hartree with a Monkhorst Pack k-point \cite{monkhorst1976special} sampling of 9x9x1 mesh. The convergence criteria of the self consistent loop is set at a value of 10$^{-5}$ Hartree.

We first confirm the band gap of WS$_2$ unit cell and find the value to be 2.07 eV, nearly consistent with the earlier reports \cite{doi:10.1021/jp300079d, kumar2012electronic}. A 5x5 supercell is formed using the optimized unit cell which is computationally appropriate to study the doping similar to the one reported here\cite{PhysRevB.88.075420}. The interface strain between the pure and doped WS$_2$ supercell with $<$111$>$ cleaved surface of gold and palladium is found to be 1.3\% and 2.1\% respectively. Doping is achieved by replacing one sulfur atom with chlorine atom in the supercell which amounts to 2\% of the total sulfur atoms. For evaluating E$_{Form}$ for sulfur substitution, X$_2$ dimers of each Cl$_2$ and S$_2$ are taken and then the total energy of the substitutional atoms and host atoms in the formation energy equation\cite{:/content/aip/journal/jap/95/8/10.1063/1.1682673} are calculated. Four layers of $<$111$>$ cleaved surface of metal are then interfaced with the WS$_2$ supercell and is found relevant to study the MoS$_2$-metal interface.\cite{gong2014unusual} To minimize false interactions between the periodic interfaced geometries, vacuum length of 20 \AA \hspace{0.1cm}is considered. A force optimization of (0.001eV/\AA) is performed for both unit cell and supercell (pure and doped) of the WS$_2$ using limited memory Broyden–Fletcher-GoldfarbShannon (LBFGS) method. The volume of the structure is allowed to change using a stress optimization of (0.001eV/\AA$^{-3}$) in both unit cell and supercell of WS$_2$.

Figure 1 (a) shows the stable atomic structure of chlorine doped 5x5 supercell of WS$_2$.  We perform formation energy calculations to find the thermodynamically stable position for sulfur substitution by chlorine atoms with E$_{Form}$ values as 2.92 eV.  This supercell is interfaced with $<$111$>$ cleaved surface of Au and Pd respectively. While interfacing the lattice constant of WS$_2$ is fixed and the metal slab is subjected to match with it. Figure 1 (c) and (d) shows the ball and stick model of the doped 5x5 WS$_2$-Au and WS$_2$-Pd interface. Total energy calculations reveal 2.7 \AA \hspace{0.1cm}and 2.2 \AA \hspace{0.1cm} as the equilibrium interlayer separation between the WS$_2$ (pure and doped) with Au and Pd metal slabs. The binding energy values obtained per sulfur atom is -0.18 eV (pure) and -0.19 eV (doped) for gold and -0.39 eV (pure) and -0.4 eV (doped) for palladium respectively. We calculate the work function of pure and doped WS$_2$ and metals Au and Pd and are shown in Figure 1(f).

Substitution by chlorine doping amounts to 2\% of total sulfur atoms which leads to a very high doping concentration in 5x5 supercell of WS$_2$. However, to realize the substitution by considering the experimental doping concentration value of 6 x 10$^{11}$ cm$^{-2}$ studied here Ref.(14), a much bigger supercell is required leading to further increase in computational burden. To reduce the computational cost and study the effect of doping on SBH reduction we limit ourselves with a 5x5 supercell.

\begin{figure}
\includegraphics[width =0.8\columnwidth]{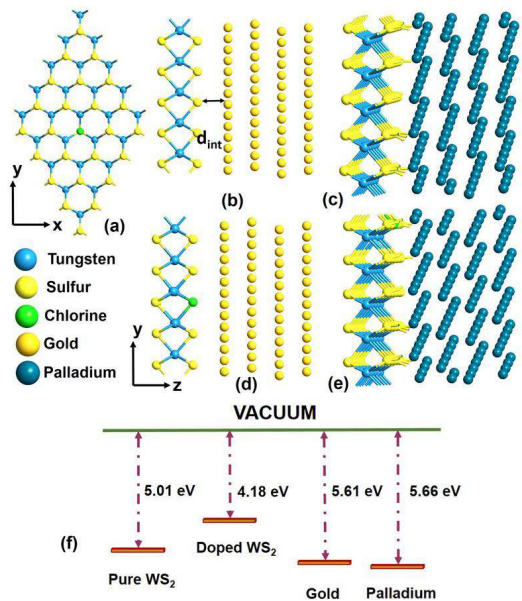}
\caption{(a) Chlorine substituted 5x5 supercell of WS$_2$ presented using the ball-stick model. Atomic model of the pure WS$_2$ 5x5 supercell (b),(c) and doped supercell (d),(e) interfaced with $<$111$>$ cleaved surface of face centered cubic (FCC) metals Gold and Palladium. The configuration is stable when the interlayer distance is 2.7 \AA \hspace{0.1cm} for gold and 2.2 \AA \hspace{0.1cm} for palladium. d$_{int}$ denotes the equilibrium distance with minimum binding energy along the z-direction. (f) Work function values of pure and doped WS$_2$ along with the metals Au and Pd calculated using DFT simulations.}
\end{figure}

The electronic structures of both pure and doped WS$_2$ along with the density of states are compared in Figure 2 to highlight the effect of doping. The band gap of pure WS$_2$ is 2.07 eV. The positions of CB and Valence Band (VB) is verified with PDOS kept along the side of bandstructure.  We see that substitution of single chlorine dopant shifts the energy bands towards the E$_F$ and creates midgap states in the vicinity of E$_F$. The number of midgap states is 2 , one of which splits into two bands between the high symmetry  points of brillouin zone shown in the inset of Figure 2 (c). The two mid gap states have significant contribution in the PDOS as well. The main contributor of these states is chlorine which is shown in inset of Figure 2 (d) with purple color. The energy level is 0.184 eV for the mid gap state near the conduction band minima (CBM). The splitting of bands leads to two energy levels 0.028 eV and 0.020 eV respectively. These defect states originate from the hybridization of the Cl 3p and the W 5d states. There exists one unpaired extra electron due to n-type doping by chlorine.  As tungsten lies in the 6th period of periodic table, chlorine, which has an extra negative charge, leads to the energy splitting of W d-orbitals and thus contribute to an extra mid gap state.\cite{}  The band gap in doped WS$_2$ is also different as compared to the pure supercell (by 0.05 eV), which is attributed to the hybridization and change in the atomic structure due to chlorine substitution. 

\begin{figure}
\includegraphics[width=0.8\columnwidth]{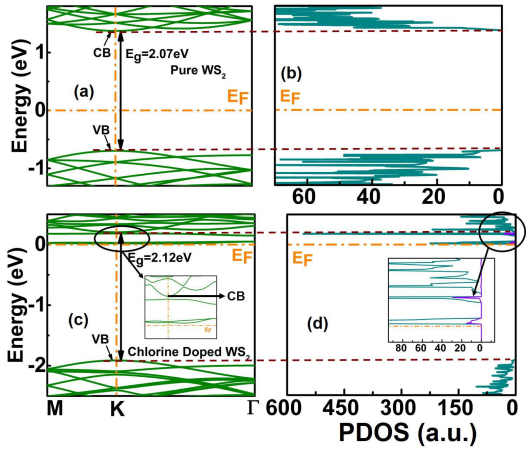}
\caption{ Projected bandstructure and DOS of (a)-(b) pure WS$_2$ and (c)-(d)chlorine doped WS$_2$. PDOS placed along the side verifies the position of CB and VB. Doping with chlorine leads to appearance of mid gap states in the WS$_2$ band gap and their contribution is studied using the DOS. The number of mid gap states is 2 for doped WS$_2$.The inset in (c) shows the exact position of CBM and the mid gap states. The purple line in (d) shows the chlorine contribution in PDOS and is highlighted in the inset. The contribution of mid gap states is confirmed by the PDOS and peaks in PDOS occurs at the same energy level as in bandstructure. CB, VB, E$_F$, and E$_g$ denotes conduction band edge, valence band edge, Fermi Level and band gap respectively. Fermi level is at zero energy. Dashed brown lines joins the CBM and VBM in bandstructure and DOS.  }%
\end{figure}
We then perform simulations to evaluate the nature of contact formed with WS$_2$ (pure and doped) with Au and Pd. The Schottky barrier height is assessed using the electronic structure and DOS of projected WS$_2$. Valence electron density and charge density difference are used to estimate the charge transfer across the interface. The localisation of electrons at the WS$_2$-metal interface  is studied using the ELF.

\begin{figure}
\includegraphics[width =0.8\columnwidth]{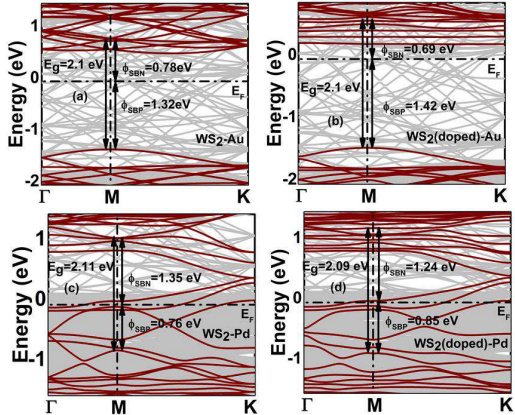}
\caption{Projected electronic bandstructure of WS$_2$-metal (Au and Pd) interface for (a)-(c) pure (b)-(d) doped supercells. The projected bandstructure (wine color lines) is superimposed on the total bandstructure WS$_2$-metal systems(grey lines). Au forms an n-type SB contact whereas Pd forms a p-type SB contact. $\phi_{SB,P}$and $\phi_{SB,N}$ are the p-type and n-type Schottky barrier heights. E$_g$ is the total band gap measured from the CBM and VBM. E$_F$ is the Fermi Level.}%
\end{figure}

Both p-type and n-type SBH of projected WS$_2$ (pure and doped) with Au and Pd are shown in Figure 3 (a)-(d). Since the interface involves two different atomic structures, the nature band gap is lost and projected band structure comprises of interface states.  The origin of these states lies in the complex electronic hybridization occurring at the WS$_2$-metal interface. Au is found to be physiosorped with WS$_2$ where as Pd is chemisorped. Au, being an s-electron metal shows less hybridization as compared to Pd which is a d-electron metal and this effect is similar to MoS$_2$-metal interface\cite{gong2014unusual}. The binding energy values quoted earlier confirm this nature. Moreover, the band gap regime also consists of less interface states in Au as compared to Pd (Figure 3 (a,c)).

For a pure WS$_2$-Au interface, the amount of complexity in the bandstructure is less, but for a doped interface it is very high. This happens because the atoms in the doped supercell already hybridize with the substitutional impurity. The difference is visible in Figure 3 (b) and (d) with respect to Figure 3 (a) and (c). It further makes the determination of band gap edges difficult. However, density of states along with the electronic structure of projected WS$_2$,  can be used to identify the CBM and valence band maxima (VBM) as demonstrated in Ref.8 The difference between the CBM and VBM of all the interfaces differs from the band gap of pure WS$_2$ by 0.04 eV. Strong interface hybridization causes emergence of more mid gap states for both the pure and doped supercells as visible in Figure 3 (c)-(d) for Pd due to its chemisorption nature with WS$_2$. We observe that Au shows a n-type SBH and Pd shows p-type SBH when interfaced with WS$_2$. The values of SBH are 0.78 eV and 0.76 eV respectively. In the complete band structure shown by grey lines, we see that the for Au the band lines are more dispersed in comparison to Pd. This indicates strong bonding of WS$_2$ with Pd. Moreover, below the Fermi level, we see a complete metallization for Pd leading to the formation of p-type contact. Here, we see a similarity in the nature of contact between WS$_2$ and WSe$_2$ while forming an interface with Au and Pd.\cite{PhysRevX.4.031005}   With the introduction of n-type impurity, the n-SBH exhibits a reduction for both metals. The n-SBH reduction attained is 0.09 eV for Au and 0.11 eV for Pd. We see the reduction in the n-SBH is higher for Pd in comparison to Au. Pd showing a p-type contact with WS$_2$ also exhibits a reduction in the n-SBH and increase in p-SBH. This validates the observation that the chlorine is effective in reducing n-SBH for both the type of metals forming p-type and n-type SBH with WS$_2$, and is equally competent for chemisorped and physiosorped metal interface with WS$_2$. A slight variance in the band gap value of WS$_2$ 2.07 (for pure WS$_2$ bandstructure) - 2.1 and 2.09 (projected WS$_2$ bandstructure) is also observed due to heterogeneous atomic interface. 

\begin{figure}
\includegraphics[width =0.8\columnwidth]{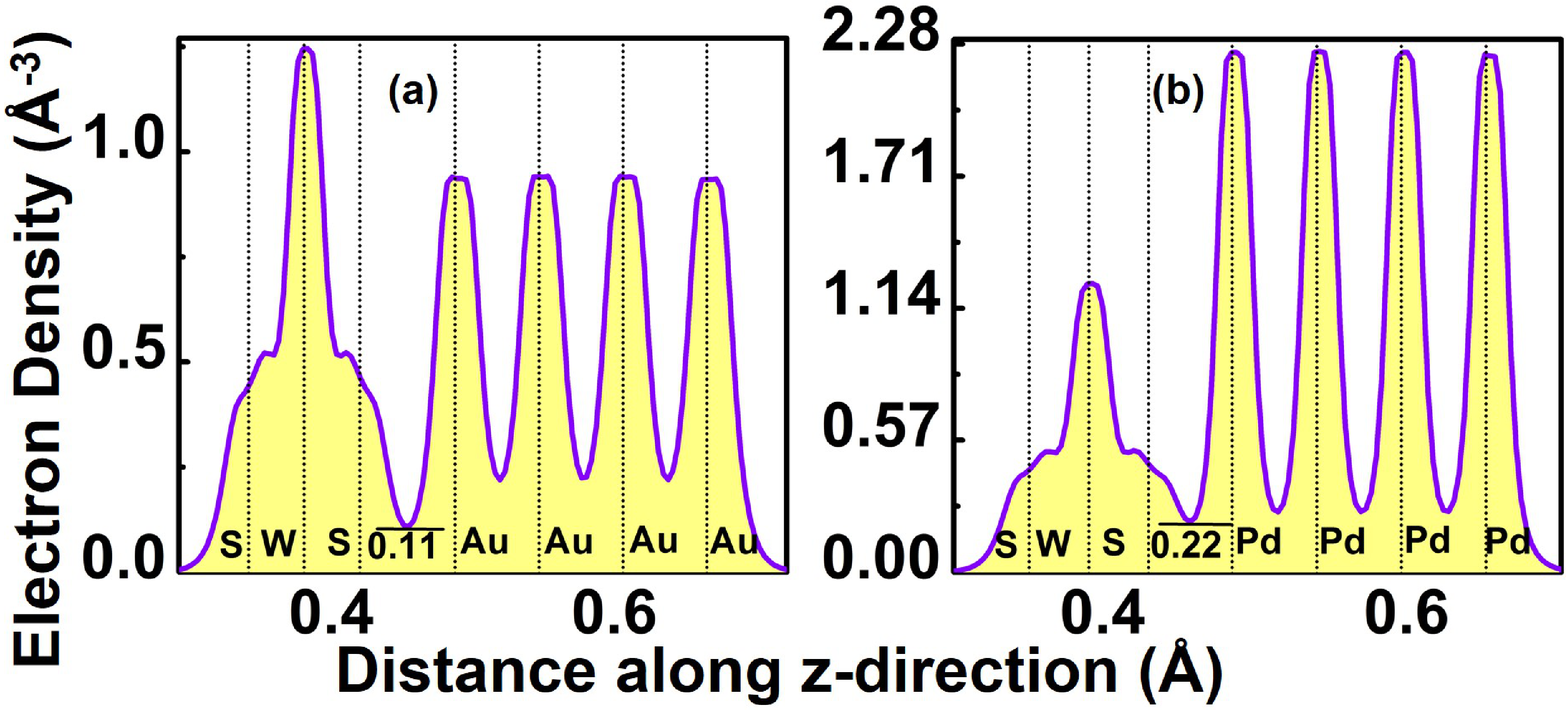}
\caption{1D projection of average valence electron density along the z-direction for pure (a) WS$_2$-Au and (b) WS$_2$-Pd interface. The value of valence electron density is evaluated between the nearest sulfur atomic layer of WS$_2$ and the nearest layer of metal formed at the WS$_2$-metal interface geometry and is shown by black horizontal lines. The position of atoms are shown by black dotted lines. }%
\end{figure}

We also evaluate the valence electron density for WS$_2$-metal interface as shown in Figure 4. The value of valence electron density is calculated as the minimum electron density at the metal-semiconductor interface. Higher values of electron density at the interface implies a better electron injection efficiency. The values with Au are  for pure 0.10727 \AA$^{-3}$ and  0.10816 \AA$^{-3}$ for doped supercell. For pure and doped supercell with Pd they are 0.21888 \AA$^{-3}$ and 0.2212 \AA$^{-3}$ respectively. The high values for Pd indicate that more charge transfer has occurred when it is interfaced with WS$_2$. Moreover the difference in the density values of doped and pure interface is higher for Pd as compared to Au.

\begin{figure}
\includegraphics[width =0.8\columnwidth]{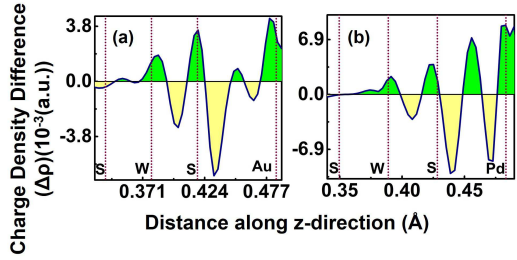}
\caption{1D projection of average charge density difference along the z-direction for pure WS$_2$ with (a) Au (b) Pd interface geometry. To study the charge transfer at the interface thoroughly, we show only single layer of gold atoms out of four. The relative positions of atoms are shown by dotted lines along the y axis. The filled green area indicates the accumulation region and the yellow indicates the depletion region. Values on the y-axis show that the charge accumulation  in Pd is nearly double as compared to Au.}%
\end{figure}
To substantiate the observations made in the above analysis, we examine the charge density difference of the respective geometries. Figure 5 shows the average charge density difference of WS$_2$-metal interface for (a) Au and (b) Pd. A dipole gets created at the interface since both accumulation and depletion regions exist. Both the regions are shown by dissimilar colors to highlight the difference between them. At the junction between the nearest sulfur atom and metal atom, two peaks exist for depletion region and one peak for accumulation region.  Moreover the accumulation shows a higher charge difference in Pd as compared to Au. A higher depletion regime implies that the charge carriers are repelled back from the surface and evinces a probability of least transmission across the interface. On the other hand, a higher accumulation regime implies that more charge is transferred across the interface. Hence we observe more hybridization for Pd rather than Au. The area calculated between the nearest S and metal atom using the methodology exhibited in Ref.8 is 0.03181 x 10$^{-3}$ for Au and 0.06031  x 10$^{-3}$ for Pd respectively and thus affirms the charge transfer is high for Pd. For pure and doped supercell-metal interface, the charge density difference curve shows a similar nature and the difference in areas for pure and doped case is higher for Pd. Because of this difference, a greater reduction on n-SBH is observed in Pd with respect to Au.

\begin{figure}
\includegraphics[width =0.8\columnwidth]{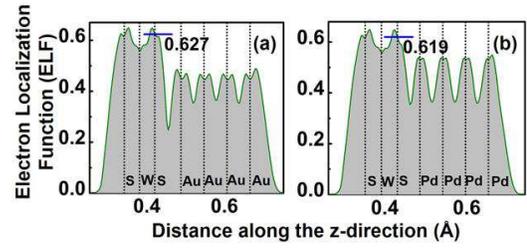}
\caption{1D projection of average electron localization function along the z-direction of pure WS$_2$ with (a) Au (b) Pd contact. The positions of atoms are shown by black dotted lines along the y-axis. The blue line indicates the value of ELF of the nearest sulfur atom in WS$_2$ monolayer.}%
\end{figure}

To study the localization of electrons at the WS$_2$-metal interface, we study ELF which is dimensionless quantity. It is defined as the possibility of finding an electron in the neighborhood of a reference electron. Its value ranges from 0$<$ELF$<$1, with the maximum limit ELF=1 corresponding to prefect localization and ELF=0.5 corresponding to electron-gas like pair probability\cite{Becke1990}. Figure 6 shows the value of ELF for WS$_2$-Au and WS$_2$-Pd interface. The value obtained for Pd is less in comparison to Au. This implies that Au is more localized than Pd indicating a better bonding for WS$_2$-Pd interface.

It is interesting to note that for TMD channel based transistors, in a top contact device structure, two types of interfaces exist. One interface is between the TMD-metal and the other is between the TMD underneath the metal contact and TMD forming the channel \citep{PhysRevX.4.031005}. Here we find the SBH formed at the first interface. For calculating the SBH at the second interface, transport simulations need to be conducted along the device length. The effective SBH combining both the interfaces can then be determined using LDOS distribution. However, carrying transport simulations employing pseudopotentials with basis functions having higher orbital contribution requires powerful computing system, as the number of atoms increase manifold from bulk-configuration to device-configuration. Nevertheless while discussing SBH and the associated contact resistance in the context of ON current of the MOSFET, the gate current is kept high and the channel is inverted. In this condition, the CBM of the TMD channel region (not touching the metal) can be assumed to be lower than CBM of the other TMD regime. Hence, the SBH evaluated in this work can be considered as the effective SBH.

In the above study we investigate the contact nature of pure and Cl doped monolayer WS$_2$ with Au$<$111$>$ and Pd$<$111$>$. First the electronic structures of pure and doped supercells are investigated. We find that band energies near to the conduction band edge align close to Fermi level with the formation of mid gap states and band splitting. Further, the pure and doped optimized supercell are adsorbed on the metals. Interface with pure supercell determines that Au is physiosorped with WS$_2$ and has an n-type SBH while Pd is chemisorped with WS$_2$ having a p-type SBH. Adsorption with Cl doped supercell shows a reduction of n-type SBH for both the metals. All the results are validated by studying charge redistribution at the WS$_2$-metal interface. The above understandings may further contribute to explore other TMD-metal interfaces, as well as in investigating various possible dopants for chalcogenide substitution.






%
%

%

\begin{acknowledgments}
 The work was supported by Science Engineering and Research Board -Department of Science and Technology (SERB DST), Govt. of India, under Grant no. SR/S3/EECE/0151/2012. The authors would like to thank Quantumwise support staff for their useful  discussions. 
\end{acknowledgments}

\bibliography{Manuscript}

\begin{thebibliography}{28}%
\makeatletter
\providecommand \@ifxundefined [1]{%
 \@ifx{#1\undefined}
}%
\providecommand \@ifnum [1]{%
 \ifnum #1\expandafter \@firstoftwo
 \else \expandafter \@secondoftwo
 \fi
}%
\providecommand \@ifx [1]{%
 \ifx #1\expandafter \@firstoftwo
 \else \expandafter \@secondoftwo
 \fi
}%
\providecommand \natexlab [1]{#1}%
\providecommand \enquote  [1]{``#1''}%
\providecommand \bibnamefont  [1]{#1}%
\providecommand \bibfnamefont [1]{#1}%
\providecommand \citenamefont [1]{#1}%
\providecommand \href@noop [0]{\@secondoftwo}%
\providecommand \href [0]{\begingroup \@sanitize@url \@href}%
\providecommand \@href[1]{\@@startlink{#1}\@@href}%
\providecommand \@@href[1]{\endgroup#1\@@endlink}%
\providecommand \@sanitize@url [0]{\catcode `\\12\catcode `\$12\catcode
  `\&12\catcode `\#12\catcode `\^12\catcode `\_12\catcode `\%12\relax}%
\providecommand \@@startlink[1]{}%
\providecommand \@@endlink[0]{}%
\providecommand \url  [0]{\begingroup\@sanitize@url \@url }%
\providecommand \@url [1]{\endgroup\@href {#1}{\urlprefix }}%
\providecommand \urlprefix  [0]{URL }%
\providecommand \Eprint [0]{\href }%
\providecommand \doibase [0]{http://dx.doi.org/}%
\providecommand \selectlanguage [0]{\@gobble}%
\providecommand \bibinfo  [0]{\@secondoftwo}%
\providecommand \bibfield  [0]{\@secondoftwo}%
\providecommand \translation [1]{[#1]}%
\providecommand \BibitemOpen [0]{}%
\providecommand \bibitemStop [0]{}%
\providecommand \bibitemNoStop [0]{.\EOS\space}%
\providecommand \EOS [0]{\spacefactor3000\relax}%
\providecommand \BibitemShut  [1]{\csname bibitem#1\endcsname}%
\let\auto@bib@innerbib\@empty
\bibitem [{\citenamefont {Radisavljevic}\ \emph {et~al.}(2011)\citenamefont
  {Radisavljevic}, \citenamefont {Radenovic}, \citenamefont {Brivio},
  \citenamefont {Giacometti},\ and\ \citenamefont
  {Kis}}]{radisavljevic2011single}%
  \BibitemOpen
  \bibfield  {author} {\bibinfo {author} {\bibfnamefont {B.}~\bibnamefont
  {Radisavljevic}}, \bibinfo {author} {\bibfnamefont {A.}~\bibnamefont
  {Radenovic}}, \bibinfo {author} {\bibfnamefont {J.}~\bibnamefont {Brivio}},
  \bibinfo {author} {\bibfnamefont {V.}~\bibnamefont {Giacometti}}, \ and\
  \bibinfo {author} {\bibfnamefont {A.}~\bibnamefont {Kis}},\ }\bibfield
  {title} {\enquote {\bibinfo {title} {Single-layer mos2 transistors},}\
  }\href@noop {} {\bibfield  {journal} {\bibinfo  {journal} {Nature
  nanotechnology}\ }\textbf {\bibinfo {volume} {6}},\ \bibinfo {pages}
  {147--150} (\bibinfo {year} {2011})}\BibitemShut {NoStop}%
\bibitem [{\citenamefont {Sik~Hwang}\ \emph {et~al.}(2012)\citenamefont
  {Sik~Hwang}, \citenamefont {Remskar}, \citenamefont {Yan}, \citenamefont
  {Protasenko}, \citenamefont {Tahy}, \citenamefont {Doo~Chae}, \citenamefont
  {Zhao}, \citenamefont {Konar}, \citenamefont {(Grace)~Xing}, \citenamefont
  {Seabaugh},\ and\ \citenamefont
  {Jena}}]{:/content/aip/journal/apl/101/1/10.1063/1.4732522}%
  \BibitemOpen
  \bibfield  {author} {\bibinfo {author} {\bibfnamefont {W.}~\bibnamefont
  {Sik~Hwang}}, \bibinfo {author} {\bibfnamefont {M.}~\bibnamefont {Remskar}},
  \bibinfo {author} {\bibfnamefont {R.}~\bibnamefont {Yan}}, \bibinfo {author}
  {\bibfnamefont {V.}~\bibnamefont {Protasenko}}, \bibinfo {author}
  {\bibfnamefont {K.}~\bibnamefont {Tahy}}, \bibinfo {author} {\bibfnamefont
  {S.}~\bibnamefont {Doo~Chae}}, \bibinfo {author} {\bibfnamefont
  {P.}~\bibnamefont {Zhao}}, \bibinfo {author} {\bibfnamefont {A.}~\bibnamefont
  {Konar}}, \bibinfo {author} {\bibfnamefont {H.}~\bibnamefont {(Grace)~Xing}},
  \bibinfo {author} {\bibfnamefont {A.}~\bibnamefont {Seabaugh}}, \ and\
  \bibinfo {author} {\bibfnamefont {D.}~\bibnamefont {Jena}},\ }\bibfield
  {title} {\enquote {\bibinfo {title} {Transistors with chemically synthesized
  layered semiconductor ws2 exhibiting 105 room temperature modulation and
  ambipolar behavior},}\ }\href@noop {} {\bibfield  {journal} {\bibinfo
  {journal} {Applied Physics Letters}\ }\textbf {\bibinfo {volume} {101}},\
  \bibinfo {eid} {013107} (\bibinfo {year} {2012})}\BibitemShut {NoStop}%
\bibitem [{\citenamefont {Liu}\ \emph {et~al.}(2014)\citenamefont {Liu},
  \citenamefont {Hu}, \citenamefont {Yue}, \citenamefont {Fera}, \citenamefont
  {Ling}, \citenamefont {Mao},\ and\ \citenamefont
  {Wei}}]{doi:10.1021/nn505253p}%
  \BibitemOpen
  \bibfield  {author} {\bibinfo {author} {\bibfnamefont {X.}~\bibnamefont
  {Liu}}, \bibinfo {author} {\bibfnamefont {J.}~\bibnamefont {Hu}}, \bibinfo
  {author} {\bibfnamefont {C.}~\bibnamefont {Yue}}, \bibinfo {author}
  {\bibfnamefont {N.~D.}\ \bibnamefont {Fera}}, \bibinfo {author}
  {\bibfnamefont {Y.}~\bibnamefont {Ling}}, \bibinfo {author} {\bibfnamefont
  {Z.}~\bibnamefont {Mao}}, \ and\ \bibinfo {author} {\bibfnamefont
  {J.}~\bibnamefont {Wei}},\ }\bibfield  {title} {\enquote {\bibinfo {title}
  {High performance field-effect transistor based on multilayer tungsten
  disulfide},}\ }\href@noop {} {\bibfield  {journal} {\bibinfo  {journal} {ACS
  Nano}\ }\textbf {\bibinfo {volume} {8}},\ \bibinfo {pages} {10396--10402}
  (\bibinfo {year} {2014})}\BibitemShut {NoStop}%
\bibitem [{\citenamefont {Das}\ and\ \citenamefont
  {Appenzeller}(2013)}]{:/content/aip/journal/apl/103/10/10.1063/1.4820408}%
  \BibitemOpen
  \bibfield  {author} {\bibinfo {author} {\bibfnamefont {S.}~\bibnamefont
  {Das}}\ and\ \bibinfo {author} {\bibfnamefont {J.}~\bibnamefont
  {Appenzeller}},\ }\bibfield  {title} {\enquote {\bibinfo {title} {Wse2 field
  effect transistors with enhanced ambipolar characteristics},}\ }\href@noop {}
  {\bibfield  {journal} {\bibinfo  {journal} {Applied Physics Letters}\
  }\textbf {\bibinfo {volume} {103}},\ \bibinfo {eid} {103501} (\bibinfo {year}
  {2013})}\BibitemShut {NoStop}%
\bibitem [{\citenamefont {Larentis}, \citenamefont {Fallahazad},\ and\
  \citenamefont
  {Tutuc}(2012)}]{:/content/aip/journal/apl/101/22/10.1063/1.4768218}%
  \BibitemOpen
  \bibfield  {author} {\bibinfo {author} {\bibfnamefont {S.}~\bibnamefont
  {Larentis}}, \bibinfo {author} {\bibfnamefont {B.}~\bibnamefont
  {Fallahazad}}, \ and\ \bibinfo {author} {\bibfnamefont {E.}~\bibnamefont
  {Tutuc}},\ }\bibfield  {title} {\enquote {\bibinfo {title} {Field-effect
  transistors and intrinsic mobility in ultra-thin mose2 layers},}\ }\href@noop
  {} {\bibfield  {journal} {\bibinfo  {journal} {Applied Physics Letters}\
  }\textbf {\bibinfo {volume} {101}},\ \bibinfo {eid} {223104} (\bibinfo {year}
  {2012})}\BibitemShut {NoStop}%
\bibitem [{\citenamefont {Pradhan}\ \emph {et~al.}(2014)\citenamefont
  {Pradhan}, \citenamefont {Rhodes}, \citenamefont {Feng}, \citenamefont {Xin},
  \citenamefont {Memaran}, \citenamefont {Moon}, \citenamefont {Terrones},
  \citenamefont {Terrones},\ and\ \citenamefont
  {Balicas}}]{doi:10.1021/nn501013c}%
  \BibitemOpen
  \bibfield  {author} {\bibinfo {author} {\bibfnamefont {N.~R.}\ \bibnamefont
  {Pradhan}}, \bibinfo {author} {\bibfnamefont {D.}~\bibnamefont {Rhodes}},
  \bibinfo {author} {\bibfnamefont {S.}~\bibnamefont {Feng}}, \bibinfo {author}
  {\bibfnamefont {Y.}~\bibnamefont {Xin}}, \bibinfo {author} {\bibfnamefont
  {S.}~\bibnamefont {Memaran}}, \bibinfo {author} {\bibfnamefont {B.-H.}\
  \bibnamefont {Moon}}, \bibinfo {author} {\bibfnamefont {H.}~\bibnamefont
  {Terrones}}, \bibinfo {author} {\bibfnamefont {M.}~\bibnamefont {Terrones}},
  \ and\ \bibinfo {author} {\bibfnamefont {L.}~\bibnamefont {Balicas}},\
  }\bibfield  {title} {\enquote {\bibinfo {title} {Field-effect transistors
  based on few-layered α-mote2},}\ }\href@noop {} {\bibfield  {journal}
  {\bibinfo  {journal} {ACS Nano}\ }\textbf {\bibinfo {volume} {8}},\ \bibinfo
  {pages} {5911--5920} (\bibinfo {year} {2014})}\BibitemShut {NoStop}%
\bibitem [{\citenamefont {Leong}\ \emph {et~al.}(2014)\citenamefont {Leong},
  \citenamefont {Luo}, \citenamefont {Li}, \citenamefont {Khoo}, \citenamefont
  {Quek},\ and\ \citenamefont {Thong}}]{leong2014low}%
  \BibitemOpen
  \bibfield  {author} {\bibinfo {author} {\bibfnamefont {W.~S.}\ \bibnamefont
  {Leong}}, \bibinfo {author} {\bibfnamefont {X.}~\bibnamefont {Luo}}, \bibinfo
  {author} {\bibfnamefont {Y.}~\bibnamefont {Li}}, \bibinfo {author}
  {\bibfnamefont {K.~H.}\ \bibnamefont {Khoo}}, \bibinfo {author}
  {\bibfnamefont {S.~Y.}\ \bibnamefont {Quek}}, \ and\ \bibinfo {author}
  {\bibfnamefont {J.~T.}\ \bibnamefont {Thong}},\ }\bibfield  {title} {\enquote
  {\bibinfo {title} {Low resistance metal contacts to mos2 devices with
  nickel-etched-graphene electrodes},}\ }\href@noop {} {\bibfield  {journal}
  {\bibinfo  {journal} {ACS nano}\ }\textbf {\bibinfo {volume} {9}},\ \bibinfo
  {pages} {869--877} (\bibinfo {year} {2014})}\BibitemShut {NoStop}%
\bibitem [{\citenamefont {Chanana}\ and\ \citenamefont
  {Mahapatra}(2015)}]{7116528}%
  \BibitemOpen
  \bibfield  {author} {\bibinfo {author} {\bibfnamefont {A.}~\bibnamefont
  {Chanana}}\ and\ \bibinfo {author} {\bibfnamefont {S.}~\bibnamefont
  {Mahapatra}},\ }\bibfield  {title} {\enquote {\bibinfo {title} {Theoretical
  insights to niobium-doped monolayer mos2 gold contact},}\ }\href@noop {}
  {\bibfield  {journal} {\bibinfo  {journal} {Electron Devices, IEEE
  Transactions on}\ }\textbf {\bibinfo {volume} {62}},\ \bibinfo {pages}
  {2346--2351} (\bibinfo {year} {2015})}\BibitemShut {NoStop}%
\bibitem [{\citenamefont {Das}\ \emph {et~al.}(2013)\citenamefont {Das},
  \citenamefont {Chen}, \citenamefont {Penumatcha},\ and\ \citenamefont
  {Appenzeller}}]{doi:10.1021/nl303583v}%
  \BibitemOpen
  \bibfield  {author} {\bibinfo {author} {\bibfnamefont {S.}~\bibnamefont
  {Das}}, \bibinfo {author} {\bibfnamefont {H.-Y.}\ \bibnamefont {Chen}},
  \bibinfo {author} {\bibfnamefont {A.~V.}\ \bibnamefont {Penumatcha}}, \ and\
  \bibinfo {author} {\bibfnamefont {J.}~\bibnamefont {Appenzeller}},\
  }\bibfield  {title} {\enquote {\bibinfo {title} {High performance multilayer
  mos2 transistors with scandium contacts},}\ }\href@noop {} {\bibfield
  {journal} {\bibinfo  {journal} {Nano Letters}\ }\textbf {\bibinfo {volume}
  {13}},\ \bibinfo {pages} {100--105} (\bibinfo {year} {2013})}\BibitemShut
  {NoStop}%
\bibitem [{\citenamefont {Kang}, \citenamefont {Liu},\ and\ \citenamefont
  {Banerjee}(2014)}]{:/content/aip/journal/apl/104/9/10.1063/1.4866340}%
  \BibitemOpen
  \bibfield  {author} {\bibinfo {author} {\bibfnamefont {J.}~\bibnamefont
  {Kang}}, \bibinfo {author} {\bibfnamefont {W.}~\bibnamefont {Liu}}, \ and\
  \bibinfo {author} {\bibfnamefont {K.}~\bibnamefont {Banerjee}},\ }\bibfield
  {title} {\enquote {\bibinfo {title} {High-performance mos2 transistors with
  low-resistance molybdenum contacts},}\ }\href@noop {} {\bibfield  {journal}
  {\bibinfo  {journal} {Applied Physics Letters}\ }\textbf {\bibinfo {volume}
  {104}},\ \bibinfo {eid} {093106} (\bibinfo {year} {2014})}\BibitemShut
  {NoStop}%
\bibitem [{\citenamefont {Laskar}\ \emph {et~al.}(2014)\citenamefont {Laskar},
  \citenamefont {Nath}, \citenamefont {Ma}, \citenamefont {Lee~II},
  \citenamefont {Lee}, \citenamefont {Kent}, \citenamefont {Yang},
  \citenamefont {Mishra}, \citenamefont {Roldan}, \citenamefont {Idrobo} \emph
  {et~al.}}]{laskar2014p}%
  \BibitemOpen
  \bibfield  {author} {\bibinfo {author} {\bibfnamefont {M.~R.}\ \bibnamefont
  {Laskar}}, \bibinfo {author} {\bibfnamefont {D.~N.}\ \bibnamefont {Nath}},
  \bibinfo {author} {\bibfnamefont {L.}~\bibnamefont {Ma}}, \bibinfo {author}
  {\bibfnamefont {E.~W.}\ \bibnamefont {Lee~II}}, \bibinfo {author}
  {\bibfnamefont {C.~H.}\ \bibnamefont {Lee}}, \bibinfo {author} {\bibfnamefont
  {T.}~\bibnamefont {Kent}}, \bibinfo {author} {\bibfnamefont {Z.}~\bibnamefont
  {Yang}}, \bibinfo {author} {\bibfnamefont {R.}~\bibnamefont {Mishra}},
  \bibinfo {author} {\bibfnamefont {M.~A.}\ \bibnamefont {Roldan}}, \bibinfo
  {author} {\bibfnamefont {J.-C.}\ \bibnamefont {Idrobo}},  \emph {et~al.},\
  }\bibfield  {title} {\enquote {\bibinfo {title} {P-type doping of mos2 thin
  films using nb},}\ }\href@noop {} {\bibfield  {journal} {\bibinfo  {journal}
  {Applied Physics Letters}\ }\textbf {\bibinfo {volume} {104}},\ \bibinfo
  {pages} {092104} (\bibinfo {year} {2014})}\BibitemShut {NoStop}%
\bibitem [{\citenamefont {Suh}\ \emph {et~al.}(2014)\citenamefont {Suh},
  \citenamefont {Park}, \citenamefont {Lin}, \citenamefont {Fu}, \citenamefont
  {Park}, \citenamefont {Jung}, \citenamefont {Chen}, \citenamefont {Ko},
  \citenamefont {Jang}, \citenamefont {Sun}, \citenamefont {Sinclair},
  \citenamefont {Chang}, \citenamefont {Tongay},\ and\ \citenamefont
  {Wu}}]{doi:10.1021/nl503251h}%
  \BibitemOpen
  \bibfield  {author} {\bibinfo {author} {\bibfnamefont {J.}~\bibnamefont
  {Suh}}, \bibinfo {author} {\bibfnamefont {T.-E.}\ \bibnamefont {Park}},
  \bibinfo {author} {\bibfnamefont {D.-Y.}\ \bibnamefont {Lin}}, \bibinfo
  {author} {\bibfnamefont {D.}~\bibnamefont {Fu}}, \bibinfo {author}
  {\bibfnamefont {J.}~\bibnamefont {Park}}, \bibinfo {author} {\bibfnamefont
  {H.~J.}\ \bibnamefont {Jung}}, \bibinfo {author} {\bibfnamefont
  {Y.}~\bibnamefont {Chen}}, \bibinfo {author} {\bibfnamefont {C.}~\bibnamefont
  {Ko}}, \bibinfo {author} {\bibfnamefont {C.}~\bibnamefont {Jang}}, \bibinfo
  {author} {\bibfnamefont {Y.}~\bibnamefont {Sun}}, \bibinfo {author}
  {\bibfnamefont {R.}~\bibnamefont {Sinclair}}, \bibinfo {author}
  {\bibfnamefont {J.}~\bibnamefont {Chang}}, \bibinfo {author} {\bibfnamefont
  {S.}~\bibnamefont {Tongay}}, \ and\ \bibinfo {author} {\bibfnamefont
  {J.}~\bibnamefont {Wu}},\ }\bibfield  {title} {\enquote {\bibinfo {title}
  {Doping against the native propensity of mos2: Degenerate hole doping by
  cation substitution},}\ }\href {\doibase 10.1021/nl503251h} {\bibfield
  {journal} {\bibinfo  {journal} {Nano Letters}\ }\textbf {\bibinfo {volume}
  {14}},\ \bibinfo {pages} {6976--6982} (\bibinfo {year} {2014})}\BibitemShut
  {NoStop}%
\bibitem [{\citenamefont {Liu}\ \emph {et~al.}(2011)\citenamefont {Liu},
  \citenamefont {Bala~Kumar}, \citenamefont {Ouyang},\ and\ \citenamefont
  {Guo}}]{5959195}%
  \BibitemOpen
  \bibfield  {author} {\bibinfo {author} {\bibfnamefont {L.}~\bibnamefont
  {Liu}}, \bibinfo {author} {\bibfnamefont {S.}~\bibnamefont {Bala~Kumar}},
  \bibinfo {author} {\bibfnamefont {Y.}~\bibnamefont {Ouyang}}, \ and\ \bibinfo
  {author} {\bibfnamefont {J.}~\bibnamefont {Guo}},\ }\bibfield  {title}
  {\enquote {\bibinfo {title} {Performance limits of monolayer transition metal
  dichalcogenide transistors},}\ }\href@noop {} {\bibfield  {journal} {\bibinfo
   {journal} {Electron Devices, IEEE Transactions on}\ }\textbf {\bibinfo
  {volume} {58}},\ \bibinfo {pages} {3042--3047} (\bibinfo {year}
  {2011})}\BibitemShut {NoStop}%
\bibitem [{\citenamefont {Yang}\ \emph {et~al.}(2014)\citenamefont {Yang},
  \citenamefont {Majumdar}, \citenamefont {Liu}, \citenamefont {Du},
  \citenamefont {Wu}, \citenamefont {Hatzistergos}, \citenamefont {Hung},
  \citenamefont {Tieckelmann}, \citenamefont {Tsai}, \citenamefont {Hobbs},\
  and\ \citenamefont {Ye}}]{doi:10.1021/nl502603d}%
  \BibitemOpen
  \bibfield  {author} {\bibinfo {author} {\bibfnamefont {L.}~\bibnamefont
  {Yang}}, \bibinfo {author} {\bibfnamefont {K.}~\bibnamefont {Majumdar}},
  \bibinfo {author} {\bibfnamefont {H.}~\bibnamefont {Liu}}, \bibinfo {author}
  {\bibfnamefont {Y.}~\bibnamefont {Du}}, \bibinfo {author} {\bibfnamefont
  {H.}~\bibnamefont {Wu}}, \bibinfo {author} {\bibfnamefont {M.}~\bibnamefont
  {Hatzistergos}}, \bibinfo {author} {\bibfnamefont {P.~Y.}\ \bibnamefont
  {Hung}}, \bibinfo {author} {\bibfnamefont {R.}~\bibnamefont {Tieckelmann}},
  \bibinfo {author} {\bibfnamefont {W.}~\bibnamefont {Tsai}}, \bibinfo {author}
  {\bibfnamefont {C.}~\bibnamefont {Hobbs}}, \ and\ \bibinfo {author}
  {\bibfnamefont {P.~D.}\ \bibnamefont {Ye}},\ }\bibfield  {title} {\enquote
  {\bibinfo {title} {Chloride molecular doping technique on 2d materials: Ws2
  and mos2},}\ }\href@noop {} {\bibfield  {journal} {\bibinfo  {journal} {Nano
  Letters}\ }\textbf {\bibinfo {volume} {14}},\ \bibinfo {pages} {6275--6280}
  (\bibinfo {year} {2014})}\BibitemShut {NoStop}%
\bibitem [{\citenamefont {Khomyakov}\ \emph {et~al.}(2009)\citenamefont
  {Khomyakov}, \citenamefont {Giovannetti}, \citenamefont {Rusu}, \citenamefont
  {Brocks}, \citenamefont {van~den Brink},\ and\ \citenamefont
  {Kelly}}]{PhysRevB.79.195425}%
  \BibitemOpen
  \bibfield  {author} {\bibinfo {author} {\bibfnamefont {P.~A.}\ \bibnamefont
  {Khomyakov}}, \bibinfo {author} {\bibfnamefont {G.}~\bibnamefont
  {Giovannetti}}, \bibinfo {author} {\bibfnamefont {P.~C.}\ \bibnamefont
  {Rusu}}, \bibinfo {author} {\bibfnamefont {G.}~\bibnamefont {Brocks}},
  \bibinfo {author} {\bibfnamefont {J.}~\bibnamefont {van~den Brink}}, \ and\
  \bibinfo {author} {\bibfnamefont {P.~J.}\ \bibnamefont {Kelly}},\ }\bibfield
  {title} {\enquote {\bibinfo {title} {First-principles study of the
  interaction and charge transfer between graphene and metals},}\ }\href@noop
  {} {\bibfield  {journal} {\bibinfo  {journal} {Phys. Rev. B}\ }\textbf
  {\bibinfo {volume} {79}},\ \bibinfo {pages} {195425} (\bibinfo {year}
  {2009})}\BibitemShut {NoStop}%
\bibitem [{\citenamefont {Gong}\ \emph {et~al.}(2010)\citenamefont {Gong},
  \citenamefont {Lee}, \citenamefont {Shan}, \citenamefont {Vogel},
  \citenamefont {Wallace},\ and\ \citenamefont
  {Cho}}]{:/content/aip/journal/jap/108/12/10.1063/1.3524232}%
  \BibitemOpen
  \bibfield  {author} {\bibinfo {author} {\bibfnamefont {C.}~\bibnamefont
  {Gong}}, \bibinfo {author} {\bibfnamefont {G.}~\bibnamefont {Lee}}, \bibinfo
  {author} {\bibfnamefont {B.}~\bibnamefont {Shan}}, \bibinfo {author}
  {\bibfnamefont {E.~M.}\ \bibnamefont {Vogel}}, \bibinfo {author}
  {\bibfnamefont {R.~M.}\ \bibnamefont {Wallace}}, \ and\ \bibinfo {author}
  {\bibfnamefont {K.}~\bibnamefont {Cho}},\ }\bibfield  {title} {\enquote
  {\bibinfo {title} {First-principles study of metal–graphene interfaces},}\
  }\href {\doibase http://dx.doi.org/10.1063/1.3524232} {\bibfield  {journal}
  {\bibinfo  {journal} {Journal of Applied Physics}\ }\textbf {\bibinfo
  {volume} {108}},\ \bibinfo {pages} {--} (\bibinfo {year} {2010})}\BibitemShut
  {NoStop}%
\bibitem [{\citenamefont {Popov}, \citenamefont {Seifert},\ and\ \citenamefont
  {Tom\'anek}(2012)}]{PhysRevLett.108.156802}%
  \BibitemOpen
  \bibfield  {author} {\bibinfo {author} {\bibfnamefont {I.}~\bibnamefont
  {Popov}}, \bibinfo {author} {\bibfnamefont {G.}~\bibnamefont {Seifert}}, \
  and\ \bibinfo {author} {\bibfnamefont {D.}~\bibnamefont {Tom\'anek}},\
  }\bibfield  {title} {\enquote {\bibinfo {title} {Designing electrical
  contacts to ${\mathrm{mos}}_{2}$ monolayers: A computational study},}\
  }\href@noop {} {\bibfield  {journal} {\bibinfo  {journal} {Phys. Rev. Lett.}\
  }\textbf {\bibinfo {volume} {108}},\ \bibinfo {pages} {156802} (\bibinfo
  {year} {2012})}\BibitemShut {NoStop}%
\bibitem [{\citenamefont {Kang}\ \emph {et~al.}(2014)\citenamefont {Kang},
  \citenamefont {Liu}, \citenamefont {Sarkar}, \citenamefont {Jena},\ and\
  \citenamefont {Banerjee}}]{PhysRevX.4.031005}%
  \BibitemOpen
  \bibfield  {author} {\bibinfo {author} {\bibfnamefont {J.}~\bibnamefont
  {Kang}}, \bibinfo {author} {\bibfnamefont {W.}~\bibnamefont {Liu}}, \bibinfo
  {author} {\bibfnamefont {D.}~\bibnamefont {Sarkar}}, \bibinfo {author}
  {\bibfnamefont {D.}~\bibnamefont {Jena}}, \ and\ \bibinfo {author}
  {\bibfnamefont {K.}~\bibnamefont {Banerjee}},\ }\bibfield  {title} {\enquote
  {\bibinfo {title} {Computational study of metal contacts to monolayer
  transition-metal dichalcogenide semiconductors},}\ }\href {\doibase
  10.1103/PhysRevX.4.031005} {\bibfield  {journal} {\bibinfo  {journal} {Phys.
  Rev. X}\ }\textbf {\bibinfo {volume} {4}},\ \bibinfo {pages} {031005}
  (\bibinfo {year} {2014})}\BibitemShut {NoStop}%
\bibitem [{Qum()}]{QumWS}%
  \BibitemOpen
  \href@noop {} {\enquote {\bibinfo {title} {{Atomistix ToolKit v.15.beta
  Quantumwise }},}\ }\bibinfo {howpublished}
  {\url{http://quantumwise.com/}}\BibitemShut {NoStop}%
\bibitem [{\citenamefont {Perdew}\ and\ \citenamefont
  {Zunger}(1981)}]{PhysRevB.23.5048}%
  \BibitemOpen
  \bibfield  {author} {\bibinfo {author} {\bibfnamefont {J.~P.}\ \bibnamefont
  {Perdew}}\ and\ \bibinfo {author} {\bibfnamefont {A.}~\bibnamefont
  {Zunger}},\ }\bibfield  {title} {\enquote {\bibinfo {title} {Self-interaction
  correction to density-functional approximations for many-electron systems},}\
  }\href {\doibase 10.1103/PhysRevB.23.5048} {\bibfield  {journal} {\bibinfo
  {journal} {Phys. Rev. B}\ }\textbf {\bibinfo {volume} {23}},\ \bibinfo
  {pages} {5048--5079} (\bibinfo {year} {1981})}\BibitemShut {NoStop}%
\bibitem [{\citenamefont {Hartwigsen}, \citenamefont {Goedecker},\ and\
  \citenamefont {Hutter}(1998)}]{PhysRevB.58.3641}%
  \BibitemOpen
  \bibfield  {author} {\bibinfo {author} {\bibfnamefont {C.}~\bibnamefont
  {Hartwigsen}}, \bibinfo {author} {\bibfnamefont {S.}~\bibnamefont
  {Goedecker}}, \ and\ \bibinfo {author} {\bibfnamefont {J.}~\bibnamefont
  {Hutter}},\ }\bibfield  {title} {\enquote {\bibinfo {title} {Relativistic
  separable dual-space gaussian pseudopotentials from h to rn},}\ }\href@noop
  {} {\bibfield  {journal} {\bibinfo  {journal} {Phys. Rev. B}\ }\textbf
  {\bibinfo {volume} {58}},\ \bibinfo {pages} {3641--3662} (\bibinfo {year}
  {1998})}\BibitemShut {NoStop}%
\bibitem [{\citenamefont {Monkhorst}\ and\ \citenamefont
  {Pack}(1976)}]{monkhorst1976special}%
  \BibitemOpen
  \bibfield  {author} {\bibinfo {author} {\bibfnamefont {H.~J.}\ \bibnamefont
  {Monkhorst}}\ and\ \bibinfo {author} {\bibfnamefont {J.~D.}\ \bibnamefont
  {Pack}},\ }\bibfield  {title} {\enquote {\bibinfo {title} {Special points for
  brillouin-zone integrations},}\ }\href@noop {} {\bibfield  {journal}
  {\bibinfo  {journal} {Physical Review B}\ }\textbf {\bibinfo {volume} {13}},\
  \bibinfo {pages} {5188} (\bibinfo {year} {1976})}\BibitemShut {NoStop}%
\bibitem [{\citenamefont {Jiang}(2012)}]{doi:10.1021/jp300079d}%
  \BibitemOpen
  \bibfield  {author} {\bibinfo {author} {\bibfnamefont {H.}~\bibnamefont
  {Jiang}},\ }\bibfield  {title} {\enquote {\bibinfo {title} {Electronic band
  structures of molybdenum and tungsten dichalcogenides by the gw approach},}\
  }\href@noop {} {\bibfield  {journal} {\bibinfo  {journal} {The Journal of
  Physical Chemistry C}\ }\textbf {\bibinfo {volume} {116}},\ \bibinfo {pages}
  {7664--7671} (\bibinfo {year} {2012})}\BibitemShut {NoStop}%
\bibitem [{\citenamefont {Kumar}\ and\ \citenamefont
  {Ahluwalia}(2012)}]{kumar2012electronic}%
  \BibitemOpen
  \bibfield  {author} {\bibinfo {author} {\bibfnamefont {A.}~\bibnamefont
  {Kumar}}\ and\ \bibinfo {author} {\bibfnamefont {P.}~\bibnamefont
  {Ahluwalia}},\ }\bibfield  {title} {\enquote {\bibinfo {title} {Electronic
  structure of transition metal dichalcogenides monolayers 1h-mx2 (m= mo, w; x=
  s, se, te) from ab-initio theory: new direct band gap semiconductors},}\
  }\href@noop {} {\bibfield  {journal} {\bibinfo  {journal} {The European
  Physical Journal B}\ }\textbf {\bibinfo {volume} {85}},\ \bibinfo {pages}
  {1--7} (\bibinfo {year} {2012})}\BibitemShut {NoStop}%
\bibitem [{\citenamefont {Dolui}\ \emph {et~al.}(2013)\citenamefont {Dolui},
  \citenamefont {Rungger}, \citenamefont {Das~Pemmaraju},\ and\ \citenamefont
  {Sanvito}}]{PhysRevB.88.075420}%
  \BibitemOpen
  \bibfield  {author} {\bibinfo {author} {\bibfnamefont {K.}~\bibnamefont
  {Dolui}}, \bibinfo {author} {\bibfnamefont {I.}~\bibnamefont {Rungger}},
  \bibinfo {author} {\bibfnamefont {C.}~\bibnamefont {Das~Pemmaraju}}, \ and\
  \bibinfo {author} {\bibfnamefont {S.}~\bibnamefont {Sanvito}},\ }\bibfield
  {title} {\enquote {\bibinfo {title} {Possible doping strategies for
  mos${}_{2}$ monolayers: An \textit{ab initio} study},}\ }\href {\doibase
  10.1103/PhysRevB.88.075420} {\bibfield  {journal} {\bibinfo  {journal} {Phys.
  Rev. B}\ }\textbf {\bibinfo {volume} {88}},\ \bibinfo {pages} {075420}
  (\bibinfo {year} {2013})}\BibitemShut {NoStop}%
\bibitem [{\citenamefont {Van~de Walle}\ and\ \citenamefont
  {Neugebauer}(2004)}]{:/content/aip/journal/jap/95/8/10.1063/1.1682673}%
  \BibitemOpen
  \bibfield  {author} {\bibinfo {author} {\bibfnamefont {C.~G.}\ \bibnamefont
  {Van~de Walle}}\ and\ \bibinfo {author} {\bibfnamefont {J.}~\bibnamefont
  {Neugebauer}},\ }\bibfield  {title} {\enquote {\bibinfo {title}
  {First-principles calculations for defects and impurities: Applications to
  iii-nitrides},}\ }\href {\doibase http://dx.doi.org/10.1063/1.1682673}
  {\bibfield  {journal} {\bibinfo  {journal} {Journal of Applied Physics}\
  }\textbf {\bibinfo {volume} {95}},\ \bibinfo {pages} {3851--3879} (\bibinfo
  {year} {2004})}\BibitemShut {NoStop}%
\bibitem [{\citenamefont {Gong}\ \emph {et~al.}(2014)\citenamefont {Gong},
  \citenamefont {Colombo}, \citenamefont {Wallace},\ and\ \citenamefont
  {Cho}}]{gong2014unusual}%
  \BibitemOpen
  \bibfield  {author} {\bibinfo {author} {\bibfnamefont {C.}~\bibnamefont
  {Gong}}, \bibinfo {author} {\bibfnamefont {L.}~\bibnamefont {Colombo}},
  \bibinfo {author} {\bibfnamefont {R.~M.}\ \bibnamefont {Wallace}}, \ and\
  \bibinfo {author} {\bibfnamefont {K.}~\bibnamefont {Cho}},\ }\bibfield
  {title} {\enquote {\bibinfo {title} {The unusual mechanism of partial fermi
  level pinning at metal--mos2 interfaces},}\ }\href@noop {} {\bibfield
  {journal} {\bibinfo  {journal} {Nano letters}\ }\textbf {\bibinfo {volume}
  {14}},\ \bibinfo {pages} {1714--1720} (\bibinfo {year} {2014})}\BibitemShut
  {NoStop}%
\bibitem [{\citenamefont {Becke}\ and\ \citenamefont
  {Edgecombe}(1990)}]{Becke1990}%
  \BibitemOpen
  \bibfield  {author} {\bibinfo {author} {\bibfnamefont {a.~D.}\ \bibnamefont
  {Becke}}\ and\ \bibinfo {author} {\bibfnamefont {K.~E.}\ \bibnamefont
  {Edgecombe}},\ }\bibfield  {title} {\enquote {\bibinfo {title} {{A simple
  measure of electron localization in atomic and molecular systems}},}\ }\href
  {\doibase 10.1063/1.458517} {\bibfield  {journal} {\bibinfo  {journal} {J.
  Chem. Phys.}\ }\textbf {\bibinfo {volume} {92}},\ \bibinfo {pages} {5397}
  (\bibinfo {year} {1990})}\BibitemShut {NoStop}%
\end{thebibliography}%

\end{document}